\documentclass[prb,twocolumn]{revtex4}

\usepackage{amsmath,amssymb,mathrsfs}
\usepackage{psfrag}
\usepackage{graphicx}
\usepackage{graphics}
\usepackage{epsfig}
\usepackage{bm}
\usepackage{color}
\usepackage{verbatim,color,ulem}
\usepackage{dblfloatfix}
\usepackage[hidelinks]{hyperref}

\newcommand{\beq}{\begin{equation}}
\newcommand{\eeq}{\end{equation}}
\newcommand{\beqa}{\begin{eqnarray}}
\newcommand{\eeqa}{\end{eqnarray}}
\usepackage{setspace}

\usepackage{mathtools}

\newcommand{\cblue}{\color{black}}

\begin{document}

\title{Finite-size effects in the two-dimensional BCS-BEC crossover}
\author{M. Lanaro$^{1,2}$, G. Bighin$^{1}$, L. Dell'Anna$^{1,2,3}$, and
  L. Salasnich$^{1,2,3,4}$}
\affiliation{
$^{1}$Dipartimento di Fisica e Astronomia ``Galileo Galilei'',  
Universit\`a di Padova, Via Marzolo 8, I-35131 Padova, Italy
 \\
$^{2}$Istituto Nazionale di Fisica Nucleare, Sezione di Padova, 
Via Marzolo 8, I-35131 Padova, Italy  
\\
$^{3}$Padua QTech Center, University of Padova, Via Gradenigo 6, I-35131 Padova, Italy
\\
$^{4}$Istituto Nazione di Ottica del Consiglio Nazionale delle Ricerche, Unita di Sesto Fiorentino, Via Carrara 2, I-50019 Sesto Fiorentino, Italy}

\date{\today}

\begin{abstract}
  We study the finite-size effects on the BCS-BEC crossover in two dimensions,
  occurring in confined fermionic superfluids. We analyze several
  thermodynamic properties, such as the chemical potential, the energy gap
  and  the superfluid density, taking into account unavoidable quantum
  fluctuations, and, by means of renormalization group procedure, we detect
  the putative Berezinskii-Kosterlitz-Thouless phase transition at finite-size.
\end{abstract}

\maketitle

\section{Introduction}

Bose and Fermi superfluids in lower spatial dimension constitute an extremely active research topic, both theoretically and experimentally. Contrary to the three-dimensional case, mean-field approaches are unreliable in two- and one-dimensional systems and the inclusion of quantum fluctuations is crucial to compare correctly the theoretical predictions with the experimental data and Monte Carlo simulations \cite{Stoof2009Ultracold,book-stringa}.  
\\In two-dimensional (2D) systems the Berezinskii-Kosterlitz-Thouless (BKT) phase transition marks the morphing from a superfluid phase characterized by quasi-condensation to a normal phase, in which vortex proliferation destroys progressively superfluidity \cite{jose40}. 
The BKT phase transition has been recently observed experimentally \cite{hadzibabic2006berezinskii,hadzibabic2021sound,jochim,enss,moritz} in Fermi gases made of alkali-metal atoms, 
and widely investigated theoretically by several groups \cite{he2022precision,ZhangBKT,TampereBKT,sademelo,bighin,bighin2017vortices}.
However an accurate study which goes beyond the mean-field level for finite-size systems is still laking.
The study of the role of the confinement for Fermi gasses is an experimentally relevant but theoretically challenging problem which requires a significant theoretical effort in order to include finite-size effects and the quantum fluctuations.
\medskip 
\\ In this work, inspired by the recent experimental achievements of box potentials \cite{box1,box2}, we investigate this puzzling problem studying a 2D Fermi gas in the crossover from the Bardeen-Cooper-Schrieffer (BCS) phase  \cite{bardeen1957microscopic} to the Bose-Einstein Condensation (BEC) phase \cite{bose1924plancks,einstein1924quantum}. First we will see how, at zero temperature, the energy $\Delta$ 
of the Cooper pairs and the chemical potential $\mu$ of the fermionic system change along the BCS-BEC crossover due to the presence of a confining
box potential. We also determine the BKT critical temperature by
calculating the superfluid fraction and, then, implementing the 
renormalization group Kosterlitz-Thouless equations \cite{bighin2017vortices},
we will take into account the finite size of the system
\cite{benfatto2009broadening,tononi2022topological,sala-bkt}. 

\section{Theoretical framework}

\begin{figure}[t]
\includegraphics[width=0.5\textwidth]{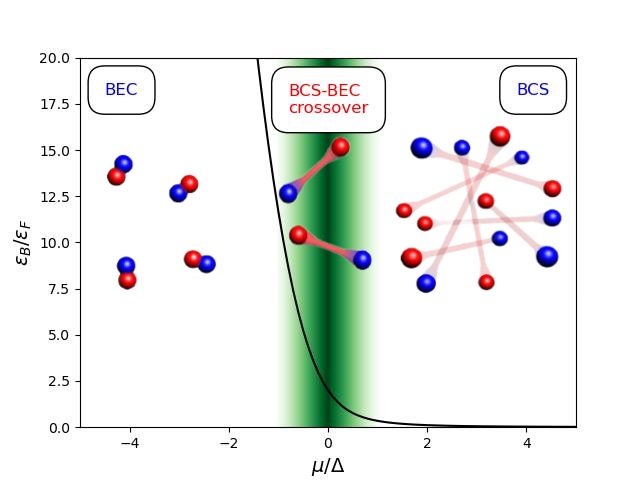}
\small 
\caption{Binding energy $\epsilon_{B}/\epsilon_{F}$ versus parameter
  $x_{0}=\mu / \Delta$. We show the regions of BCS ($x_{0} \xrightarrow{}
  \infty$) and BEC ($x_{0} \xrightarrow{}- \infty$). The green region points
  the relevant values for the BCS-BEC crossover.} 
\label{epsilon_B_explained}
\end{figure}

We will use the binding energy $\epsilon_{B}$ of Cooper pairs as tunable
interaction parameter along the BCS-BEC crossover.
Remarkably, $\epsilon_B$ depends on the ratio $\mu / \Delta$ 
by the following simple {\cblue mean-field} formula \cite{marini1998evolution}, 
{\cblue 
\begin{equation}
    \frac{\epsilon_B}{\epsilon_F} = 2 
   \frac
   {\sqrt{1 + \left({\mu\over \Delta}\right)^2} - {\mu\over \Delta}}
   {\sqrt{1 + \left({\mu\over \Delta}\right)^2} + {\mu\over\Delta}}  
    \label{perchemai}
\end{equation}
where} $\epsilon_{F}=\pi\hbar^{2} n/m$ is the 2D Fermi energy of
non-interacting two-spin-component fermions with number density $n$ and
individual mass $m$. 
This expression is a mean-field result which, nevertheless, works also beyond
the mean-field level, including Gaussian fluctuations
\cite{bighin,bighin2017vortices}. As shown in Fig. \ref{epsilon_B_explained}, 
the ratio $\mu/\Delta$ ranges from $-\infty$, when $\epsilon_B\to +\infty$
(BEC regime), to $+\infty$, when $\epsilon_B\to 0$ (BCS regime). 
Thus, contrary to the 3D case, in the 2D problem the Cooper pairs always
form a bound state. In Fig. \ref{epsilon_B_explained} we show how the
binding energy depends on $\mu/\Delta$.
The crossover regime (green region) is approximately chosen for $\mu/\Delta$
in the range $(-1,1)$ since at $\mu=0$ the Fermi surface is lost.  

{\cblue Finite-size effects are included through an infrared cutoff
  $k_{min}$ in the wavenumber $k$ of the quantum particle of mass $m$,
  which corresponds to a wavelength $L=2\pi/k_{min}$. 
  We are considering the box confinement
  such that $\epsilon_{min}=\hbar^2k_{min}/(2m)$ is the lowest
single-particle energy of the non-interacting Schr\"odinger
problem.} We will analyze the BCS-BEC crossover in two dimensions going
beyond the mean-field analysis
(both mean-field calculations and details of the beyond-mean-field approach
are reported
in the Appendices A and B, respectively).
The putative Berezinskii-Kosterlitz-Thouless phase transition
for finite systems will be obtained by means of the renormalization
group (RG) procedure. 
\begin{figure}
    \includegraphics[width=0.48\textwidth]{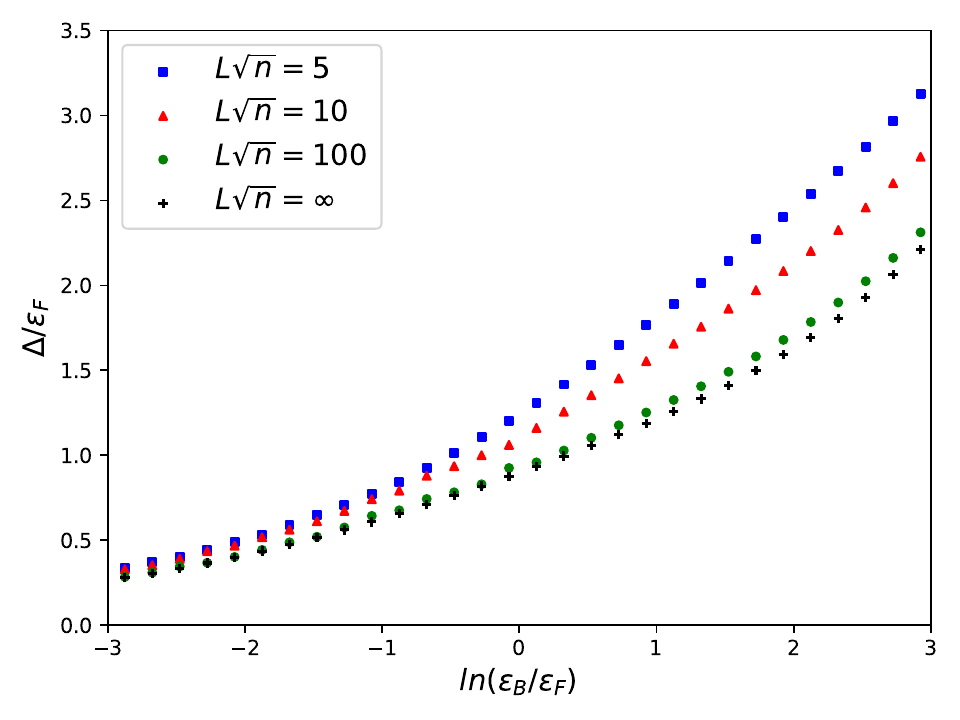}
    \hfill
    \includegraphics[width=0.48\textwidth]{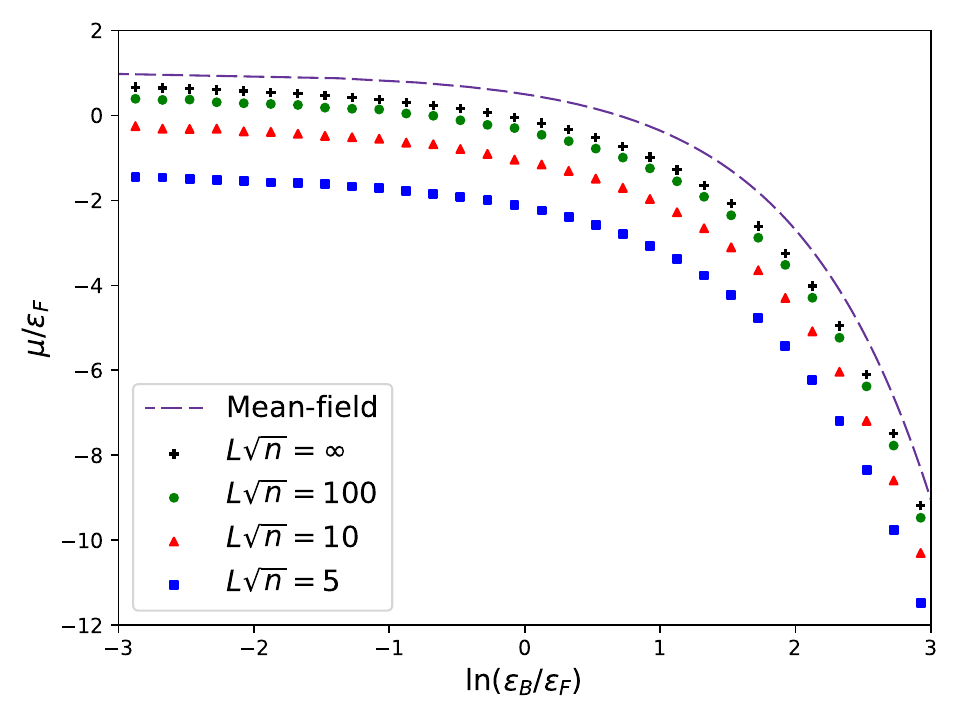}
    \small
    \caption{In the higher panel we show
      $\Delta / \epsilon_{F}$ versus $\epsilon_{B} / \epsilon_{F}$, while
      in the lower $\mu / \epsilon_{F}$ versus $\epsilon_{B} / \epsilon_{F}$
      for different values of the cutoff, respectively $L\sqrt{n}$=5
      (blue squared points), 10 (red triangled points), 100
      (green rounded points) and $\infty$ (black plus points).
      Here {\cblue $L=2\pi/k_{min}$ with $k_{min}$ the infrared cutoff}  
      while $\sqrt{n} = 1 / d$, with $d$ inter-particle distance.
      In this way, $L\sqrt{n}$ is an adimensional quantity. {\cblue
        In the lower panel we have also included the mean-field chemical
        potential $\mu$ which, contrary to the energy gap $\Delta$, 
        is size independent (see Appendix A).}}
    \label{Delta_BMF_cutoff}
\end{figure}

\section{Beyond Mean Field analysis}

The role of fluctuations is crucial in low dimensions. We aim at investigating how some quantities, such as the chemical potential and the energy gap, change in the presence of Gaussian fluctuations. 
From the gap equation, in presence of a cutoff, we get (see Appendix B) 
\begin{equation}
\Delta = \sqrt{\epsilon_{B}^{2} + 2 (\epsilon_{min} + \mu) \epsilon_{B}}
\end{equation} 
where $\epsilon_{min}=\hbar^2k_{min}^2/(2m)$, 
$\mu$ is obtained by solving the number equation at the mean-field level supplemented by the Gaussian contribution, 
$-\frac{d\Omega_{\rm GF}(\mu)}{d\mu}$. 
$\Omega_{\rm GF}$ is the beyond-mean-field gran potential which takes the form 
\begin{equation}\label{gran_pot_eq}
    \Omega_{\rm GF}=\frac{1}{2\beta} \sum_{\textbf{q},m} \ln{(\text{det } \mathbb{M}(\textbf{q}, \text{i}\Omega_{m}))}
\end{equation}
where the matrix $\mathbb{M}(\textbf{q}, \text{i}\Omega_{m})$ 
is the inverse pair fluctuation propagator, reported in Appendix B.  
Following the approach in Ref. \cite{he2015quantum}, we, then, calculate $\mu$ 
looking at the maximum of the energy density $E(\mu)=\Omega(\mu)+\mu n$,  at a given density $n$, where $\Omega(\mu)=\Omega_{MF}+\Omega_{GF}$, namely it is composed by the mean-field term and the contribution coming from the Gaussian fluctuations. 
This procedure for determining the chemical potential beyond the
mean-field approach at finite size is computationally demanding and requires
a big computational effort \cite{cloud} and polynomial fit to 
smooth the {\cblue slightly} noisy data {\cblue for the determination
  of the Berezinskii-Kosterlitz-Thouless critical temperature
  (see next Section).}
{\cblue The evaluation of the grand potential in Eq.\eqref{gran_pot_eq} requires four different integrations: two of them appear explicitly in Eq.\eqref{gran_pot_eq}, i.e. the momentum integration and the Matsubara frequency sum which can be converted to an integration in the zero-temperature limit, while two are contained in the M matrix elements, i.e. the radial and angular part of the momentum k in Eqs.\eqref{first_matrix} and \eqref{second_matrix}. The innermost two integrations are evaluated using an adaptive algorithm \cite{MultiDimIntegration}, recursively dividing the grid until the a certain relative error is required, in this case we set the maximum relative error to $10^{-5}$.
  The outermost two integrations are performed on a $200\times 200$ grid, using
  the Gauss–Kronrod quadrature formula. We verify the convergence using
  a much finer grid for selected data points.}

Finally, let us consider the superfluid density. In order to go beyond the mean-field analysis we have to consider also the collective excitations, so that the superfluid density reads 
\begin{equation}
n_{s}=n-n_{sp}-n_{col}
\end{equation}
where $n$ is the total density, $n_{sp}$ is the mean-field fermionic single particle contribution to the normal density
while $n_{col}$ is the beyond-mean-field contribution coming from the bosonic collective modes
\begin{equation}
\label{nc}
n_{col}= \frac{\beta}{2}\int_{k_m}^\infty \frac{d^2\bf{k}}{(2\pi)^2} \, {\hbar^2 k^2}{m}\frac{\exp[\beta {\cal E}_{col}({\bf k})]}{\left(\exp[\beta{\cal E}_{col}({\bf k})]-1\right)^2}
\end{equation}
If we make the following ansatz 
\begin{equation}
{\cal E}_{col}({\bf k})\simeq \hbar c_s k
\end{equation}
where $c_s$ is the sound velocity, which in its turn, depends on $\epsilon_B$, 
we can calculate Eq.~(\ref{nc}) and, defining $x_{min}=\beta \hbar c_s k_{min}$,
we get
\begin{eqnarray}
  \nonumber n_{col}=\frac{m}{4\pi \hbar^2 c_s^4}\frac{1}{\beta^3}
  \left[6\,\textrm{Li}_3(e^{x_{min}})-6x_{min}\,\textrm{Li}_2(e^{x_{min}})\right.\\
    \left. -x_{min}^2\left(\frac{x\,e^{x_{min}}}{1-e^{x_{min}}}
    +3\ln[1-e^{x_{min}}]\right)\right]
\end{eqnarray}
In the limit $x_{min}\rightarrow 0$, namely for $k_{min}\rightarrow 0$,
for an infinite system, only the first term survives and we get
\begin{equation}
\label{nc0}
n_{col}=\frac{3m}{2\pi \hbar^2 c_s^4}\frac{\zeta(3)}{\beta^3}
\end{equation}
which is the maximum value  for $n_{col}$. We see, therefore, that the collective excitations further contribute to suppressing the superfluid density. However such a suppression is less effective for finite size systems and, therefore, the superfluid density is expected to be enhanced by finite-size effects. 

\section{Berezinskii-Kosterlitz-Thouless phase transition}

From the Mermin-Wagner-Hohenberg theorem \cite{hohenberg1967existence, mermin1966absence} we know that in two dimensions the critical temperature at which we have the superconductive phase transition is $T_{c}=0$ \cite{Altland2006CondensedMF}. However, it is possible to recognize a quasi-condensate phase and the temperature under which we can see this type of behaviour is called Berezinskii-Kosterlitz-Thouless temperature ($T_{BKT}$).
\\Another important feature characterizing this phase transition is the universal jump of the superfluid density
\begin{equation}
	n_{s}(T^{-}_{BKT}) \neq n_{s}(T^{+}_{BKT}) = 0
\end{equation}
that is due to the abrupt unbinding of two vortices that typically describes this transition.
Kosterlitz derived a set of renormalized group (RG) equations able to describe the flow of vortices in the two dimensional Berezinskii-Kosterlitz-Thouless phase transition \cite{kosterlitz1974critical}. The RG equations are the following
\begin{equation}\label{KN_RG_equations}
  \left\{\begin{array}{@{}l@{}}
    \frac{d}{dl}K^{-1}_{l}(T)=4\pi^{3}y^{2}_{l}(T)\\
    \\
    \frac{d}{dl}y_{l}(T)= \left[2-\pi K_{l}(T)\right] \;y_{l}(T)
  \end{array}\right.\
\end{equation}
where $l=\ln{\left(r/d \right)}$, $r$ is the size of the superfluid system, characterized by the presence of vortices, and $d$ can be identified with the inter-particle distance.
\begin{figure}
    \includegraphics[width=0.48\textwidth]{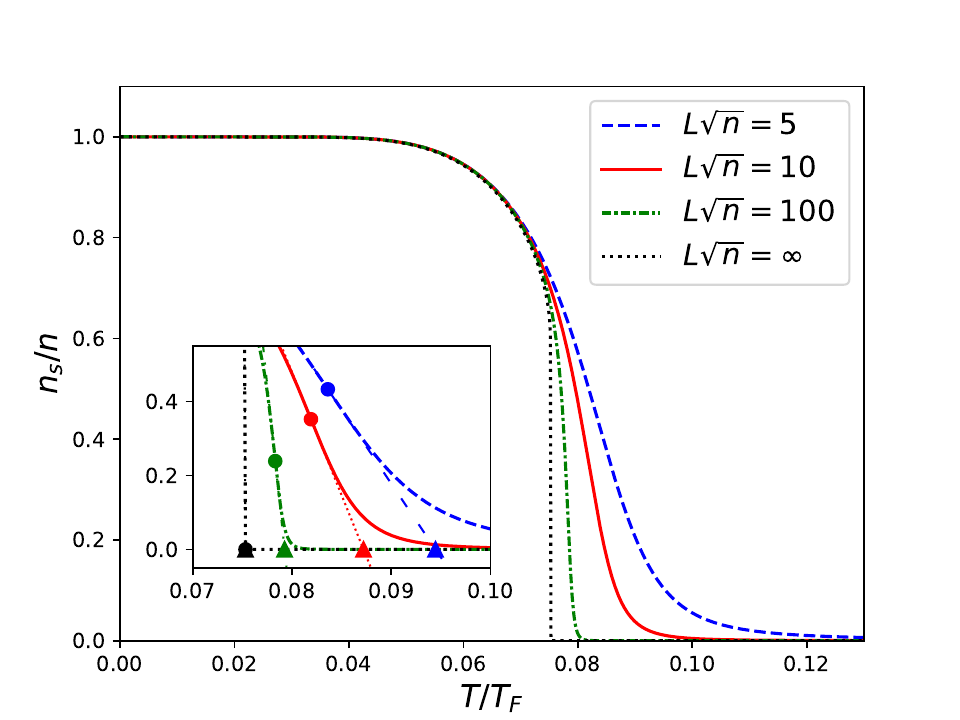}
    \small
    \caption{(Main plot) The superfluid density $n_{s}/n$ versus $T/T_{F}$ at fixed value of $\ln{\left(\epsilon_{B}/\epsilon_{F}\right)}=0.92$, for $L\sqrt{n}$=5 (blue dashed line), 10 (red solid line), 100 (green dashed-dotted line) and $\infty$ (black dotted line).
    (Inset) Zoom of the main plot. The inflection points of the superfluid density are depicted by round points while the triangular points are placed at the intersections of the slope at the inflection points with the $T$-axis.}
  \label{n_s_REN_cutoff_ln5}
\end{figure}
\begin{figure}
\includegraphics[width=0.48\textwidth]{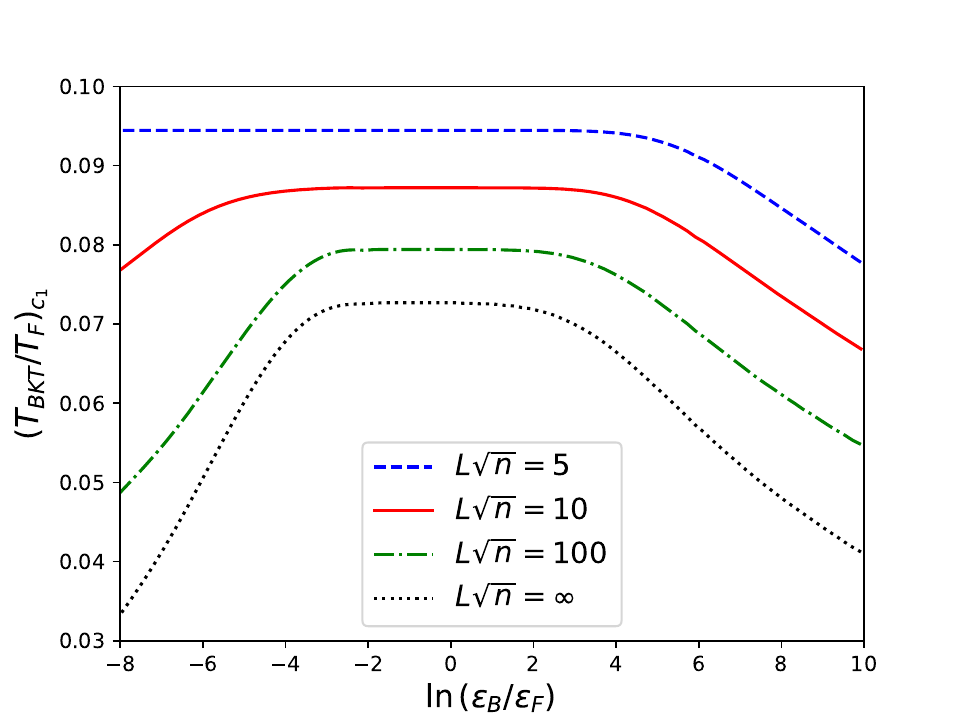}
    \small
  \caption{The critical temperature obtained through the intersection between the tangent to the superfluid density at the inflection point and the horizontal line $n_{s}/n=0$ (the triangular points in Fig. \ref{n_s_REN_cutoff_ln5}) i.e.  $(T_{BKT}/T_{F})_{c_{1}}$ against $\ln{(\epsilon_{B}/\epsilon_{F})}$ (see the main text). } \label{T_BKT_RG_cutoff_comparison}
\end{figure}

An important quantity describing the system is the phase stiffness. A fermionic superfluid can be described by the effective Hamiltonian of two dimensional XY model, which reads 
\begin{equation}
H = \frac{J}{2} \int d^{2}\textbf{r} (\nabla \theta(\textbf{r}))^{2}, 
\end{equation}
where the phase stiffness $J$ is a function of the fermion-fermion attractive strength and of the temperature \cite{bighin2017vortices}. This quantity is related to the superfluid density via \cite{fisher1973helicity}
\begin{equation}\label{4_23}
    J = \frac{\hbar^{2}}{4m} n_{s}
\end{equation}
and the superfluid density $n_{s}$ found through the Beyond Mean Field analysis must be computed with the gap energies $\Delta$ in presence of the cutoff.
From the phase stiffness, we define, then, the following quantity
\begin{equation}
    K = \frac{J}{k_{B}T} 
\end{equation}
The quantity $y$, instead, is related to the the core vortex energy $\mu_{c}$ such that
\begin{equation}
    y = \exp{\left[- \frac{\mu_{c}}{k_{B}T} \right]} 
\end{equation}
Higher order terms \cite{amit1984field}, slightly modify Eqs. (\ref{KN_RG_equations}) therefore can be neglected. 
In the framework of the two-dimensional XY model the vortex core energy is commonly expressed as
  $\mu_{c} = \frac{\pi^{2}}{4}J$, 
within the Ginzburg-Landau theory of superconducting films \cite{minnhagen1985charge}. \\
The starting point for the RG flow is determined by the bare superfluid density, calculated at a fixed temperature $T$, which defines the initial superfluid stifness $J_0(T)$. The initial conditions for the RG equations are, therefore,
\begin{equation}
  \left\{\begin{array}{@{}l@{}}
    K_{0}(T) = \frac{J_{0}(T)}{k_{B}T}\\
    \\
    y_{0}(T) = \exp{\left[ -\frac{\pi^{2}}{4} \frac{J_{0}(T)}{k_{B}T} \right]}
  \end{array}\right.\
\end{equation}

\section{Superfluid density and critical temperature} 

We must notice that also the adimensional scaling parameter $l$ of the RG
equations will be influenced by the cutoff \cite{benfatto2009broadening}. 
The inter-particle distance $d$ is taken as the square-root of the total
density of particles $n$, i.e. $d =1/\sqrt{n}$, 
while $L$ is the size of our system imposed
by a confining potential. Therefore,
the maximum value of $l$ is 
\beq
l_{max}=\ln{\left( L / d \right) } = \ln{\left( L \sqrt{n} \right)}
\eeq
In Fig. \ref{n_s_REN_cutoff_ln5} we show how the inclusion of the cutoff modifies the behavior of $n_{s}/n$ as a function of $T/T_{F}$. The main feature is the disappearance of the universal jump at the critical temperature, replaced by a curve that becomes smoother and smoother upon decreasing the size of the system.\\
Our aim is, now, to obtain the critical temperature $T_{BKT}/T_{F}$ as a function of the normalized binding energy $\epsilon_{B}/\epsilon_{F}$. 
In the limit of infinite size, the slope of the tangent of the superfluid density right at the critical temperature goes to infinity. 
For finite-size systems, instead, the superfluid density is a smooth function. The criterion for identifying the critical temperature, however, can be still based of the maximum value of the slope of the tangent of the superfluid density.
The criterion has a certain degree of arbitrariness 
therefore we can decide to follow two different procedures. In the first one, we  focus on the intersection between the tangent of $n_{s}/n$ at the inflection point, where the absolute value of the slope is maximum, and the temperature-axis, identifying this intersection as the critical temperature for a specific value of $\epsilon_{B}/\epsilon_{F}$.  As shown in Fig. \ref{n_s_REN_cutoff_ln5}, the critical temperatures obtained by this procedure are represented by the triangular points. 
The slopes are the straight lines and we call these critical temperatures $(T_{BKT}/T_{F})_{c_{1}}$.
 In this way we obtained $T_{BKT}/T_{F}$  as a function of $\epsilon_{B}/\epsilon_{F}$ for different values of $L \sqrt{n}$, shown in Fig. \ref{T_BKT_RG_cutoff_comparison}. 
 In the second procedure, also adopted in Ref. \cite{tononi2022topological}, we identify the critical temperatures as the x-coordinates of the inflection points, where $n_{s}/n$ has the maximum absolute value of the slope. In Fig. \ref{n_s_REN_cutoff_ln5} these points are the round marked ones. The plots of the critical temperature obtained by this latter procedure, called $(T_{BKT}/T_{F})_{c_{2}}$, 
 as a function of the binding energy,  for different values of $L \sqrt{n}$,  are showed in Appendix C.

\section{Conclusions}
 
The main objective of this work has been to understand how the size of the system, finite in feasible experiments, 
affects the BCS-BEC crossover in two dimensions. 
The finite-size has been introduced through an infrared cutoff in momentum space. Setting a minimum value to the wave-vector corresponds, indeed, to setting a maximum value to the wave-length, which is the size of the system. 
We have analyzed the effects of the finite size in several thermodynamic properties, such as the chemical potential, the energy gap and the superfluid density, going beyond the mean-field level by including Gaussian quantum fluctuations. Finally, by means of a renormalization group analysis and defining a couple of criteria useful to identify an effective critical temperature tailored for finite sizes, we have characterized the Berezinskii-Kosterlitz-Thouless phase transition in such systems. We believe that in the near future our theoretical predictions could be compared with the available experimental setups of strongly confined gases of fermionic atoms in quasi-two dimensional geometry. In particular, the behavior of the measured superfluid density and critical temperature will 
highlight the reliability and accuracy of
the beyond-mean-field theoretical method we have proposed. 

\section*{Acknowledgements}

CloudVeneto is acknowledged for the use of computing and storage facilities. LD and LS are partially supported by “Iniziativa Specifica Quantum” of INFN and by the Project "Frontiere Quantistiche" (Dipartimenti di Eccellenza) of the Italian Ministry for Universities and Research. ML and LS are respectively fully and partially supported by the European Quantum Flagship Project "PASQuanS2". LD and LS are partially supported by the European Union-NextGenerationEU within the National Center for HPC, Big Data and Quantum Computing [Spoke 10: Quantum Computing]. LS is partially supported by the BIRD Project "Ultracold atoms in curved geometries" of the University of Padova and by PRIN Project "Quantum Atomic Mixtures: Droplets, Topological Structures, and Vortices". LD acknowledges support from the BIRD Project n.211534 "Correlations, dynamics and topology in long-range quantum systems" granted by the University of Padova.

{\cblue 

\section*{Appendix A: Mean Field Approach} 

We will consider in what follows the mean-field analysis.
We can start with the bound state equation of the two body problem, that,
in two dimensions, takes the form
\begin{equation}\label{2D_bstate}
    -\frac{1}{g}= \frac{1}{\Omega} \sum_{ \substack{ {\bf k} \\ |{\bf k}|\geq k_{min}}} \frac{1}{\frac{\hbar^{2}k^{2}}{m}+\epsilon_{B}} 
\end{equation} 
where $m$ is the mass of a fermionic particle and $\Omega$ 
is the two-dimensional finite-size volume. The coupling constant $g$ is,
instead, the strength of the attractive potential. On the same time the
gap equation reads  
\begin{equation}\label{2D_gap}
    -\frac{1}{g}=\frac{1}{\Omega} \sum_{ \substack{ {\bf k} \\ |{\bf k}|\geq k_{min}} } \frac{1}{2\sqrt{\big(\frac{\hbar^{2}k^{2}}{2m}-\mu\big)^{2}+\Delta^{2}}} 
\end{equation}
Both Eqs.~(\ref{2D_bstate}) and (\ref{2D_gap}) have ultraviolet divergencies 
which can be regularized by subtracting one to the other
\begin{equation}\label{2D_reg_gap}
   0 = \sum_{ \substack{ {\bf k} \\ |{\bf k}|\geq k_{min}} } \left(\frac{1}{\frac{\hbar^{2}k^{2}}{m}+\epsilon_{B}}-\frac{1}{2\sqrt{\big(\frac{\hbar^{2}k^{2}}{2m}-\mu\big)^{2}+\Delta^{2}}}\right)
\end{equation}
Working with the two-dimensional continuum limit, 
$\sum_{\textbf{k}} \xrightarrow{}\frac{\Omega}{(2 \pi)^{2}}\int d^{2}\textbf{k} = \frac{\Omega}{2 \pi} \int dk \;k$
after some manipulations, we obtain the regularized gap equation in presence of an infrared cutoff
\begin{equation}\label{Gap_no_resc}
\epsilon_{B} = \sqrt{(\epsilon_{min}-\mu)^{2}+\Delta^ {2}} - (\epsilon_{min}+\mu) 
\end{equation}
where the minimum energy is
\begin{equation}
    \epsilon_{min} = \frac{\hbar^{2} k_{min}^{2}}{2m}
\end{equation}
We notice that if we impose $\epsilon_{min}=0$ we recover the gap
equation for an infinite system. The continuum limit
(semiclassical approximation) with a cutoff $k_{min}$ washes out
the geometry of the confining box, which is encoded in the
discrete energy levels of the single-particle problem.
However, the aim of our paper is to show finite-size effects
in a quite general framework.
We can now express, in the BCS limit, the number of fermionic particles as
\begin{align}
  n &= \frac{N}{\Omega} = \frac{2}{\Omega} \sum_{ \substack{ {\bf k} \\ |{\bf k}|\geq k_{min}} }\frac{1}{2}\left[ 1 - \frac{\big( \frac{\hbar^{2}k^{2}}{2m}-\mu \big)}{\sqrt{\big( \frac{\hbar^{2}k^{2}}{2m}-\mu \big)^{2} + \Delta^{2}}}\right] \label{2d_number}
   \\\nonumber 
   &= \frac{m}{2\pi \hbar^{2}} \left[ \sqrt{(\epsilon_{min}-\mu)^{2}+\Delta^{2}}-\epsilon_{min}+\mu\right] \label{2d_number_first}
\end{align}
If we consider a 2D finite-size system of $N$ fermionic particles of mass $m$ inside a 2D volume $\Omega$ and we assume two possible spin states we have
\begin{equation}\label{n_kminnneq0}
    N = 2 \sum_{ \substack{ {\bf k} \\ |{\bf k}|\geq k_{min}} } \Theta\left( \tilde\epsilon_{F} - \frac{\hbar^{2}k^{2}}{2m}\right)
\end{equation}
where $\Theta$ is the Heaviside function and ${\tilde\epsilon_{F}}$ is the Fermi energy of non-interacting fermions.
The Fermi wave number is related to the Fermi energy by the law, 
    $k_{F} = \sqrt{\frac{2m \tilde\epsilon_{F}}{\hbar^{2}}}$.
From Eq. (\ref{n_kminnneq0}) we get
\begin{equation}\label{n_kminneq0}
\tilde\epsilon_{F}= \pi \frac{\hbar^{2}}{m}n + \epsilon_{min} 
\end{equation}
As expected, for $\epsilon_{min}\xrightarrow{}0$ we get the familiar 2D Fermi energy $\tilde\epsilon_{F}=\pi\hbar^{2}n/m$ of non-interacting fermions, which is valid in the thermodynamic limit. 
We rewrite then Eq.~(\ref{n_kminneq0}) as
\begin{equation}\label{total_density_cutoff}
    n=\frac{m}{\hbar^{2}\pi}(\Tilde{\epsilon_{F}}-\epsilon_{min})
\end{equation}
If we insert the last expression into the left hand side of (\ref{2d_number})
and, after some manipulations, we obtain the following form for the number
equation of a 2D gas
\begin{equation}\label{numb_nosc}
    2\,\tilde\epsilon_{F}= \sqrt{(\epsilon_{min}-\mu)^{2}+\Delta^{2}}+\epsilon_{min}+\mu  
\end{equation}
We try to find, now, an expression for $\mu$ and $\Delta$ in terms of the binding energy $\epsilon_{B}$. 
From Eqs. (\ref{Gap_no_resc}) and (\ref{numb_nosc}) we get
\begin{equation}\label{2d_mu_eb}
   \mu= \tilde\epsilon_{F}- \epsilon_{min} - \frac{\epsilon_{B}}{2} 
\end{equation}
From Eq.~(\ref{numb_nosc}), and using Eq.~(\ref{2d_mu_eb}) we get
\begin{equation}\label{2d_delta_eb}
    \Delta = \sqrt{(2\epsilon_{B}+4\epsilon_{min})(\tilde\epsilon_{F}-\epsilon_{min})}
\end{equation}
In both cases, by imposing $\epsilon_{min}=0$ we recover the infinite-size results.
\\In order to have a consistent analysis for different values of $\epsilon_{min}$ we must look for quantities that are normalized with respect something that does not depend on $\epsilon_{min}$ itself. For this reason, we define 
\begin{equation}
    \epsilon_{F} \equiv \tilde\epsilon_{F}-\epsilon_{min}
\end{equation}
\\We can, then, rewrite (\ref{2d_mu_eb}) and (\ref{2d_delta_eb}) as
\begin{align}
      \frac{\mu}{\epsilon_{F}} &= 1-\frac{\epsilon_{B}}{2\epsilon_{F}}  \label{tildemukminneq0}
    \\ \frac{\Delta}{\epsilon_{F}} &= \sqrt{2\frac{\epsilon_{B}}{\epsilon_{F}}+4\frac{\epsilon_{min}}{\epsilon_{F}}}  \label{tildedeltakminneq0}
\end{align}
\begin{figure}
    \includegraphics[width=0.48\textwidth]{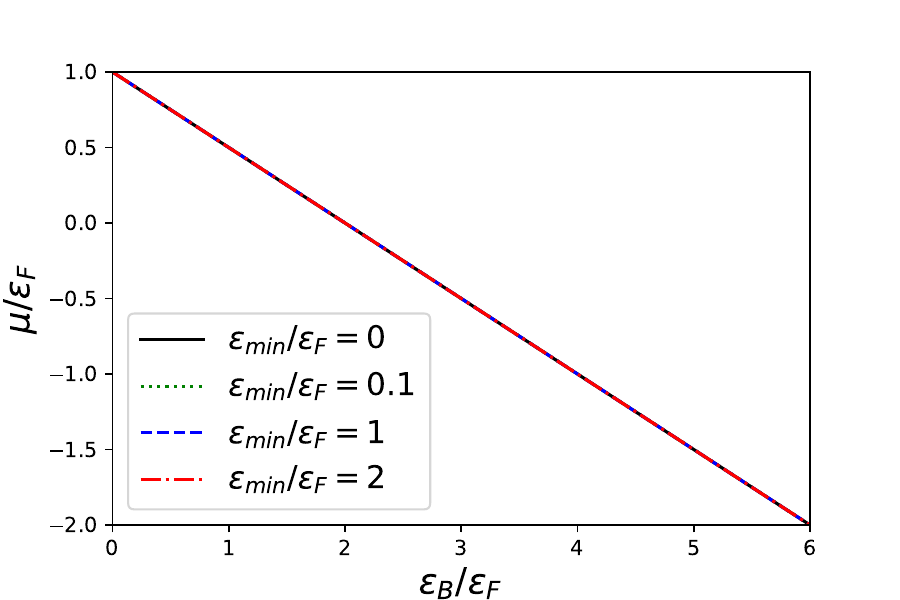}
    \label{cf_mu_MF}
  \hfill
    \includegraphics[width=0.48\textwidth]{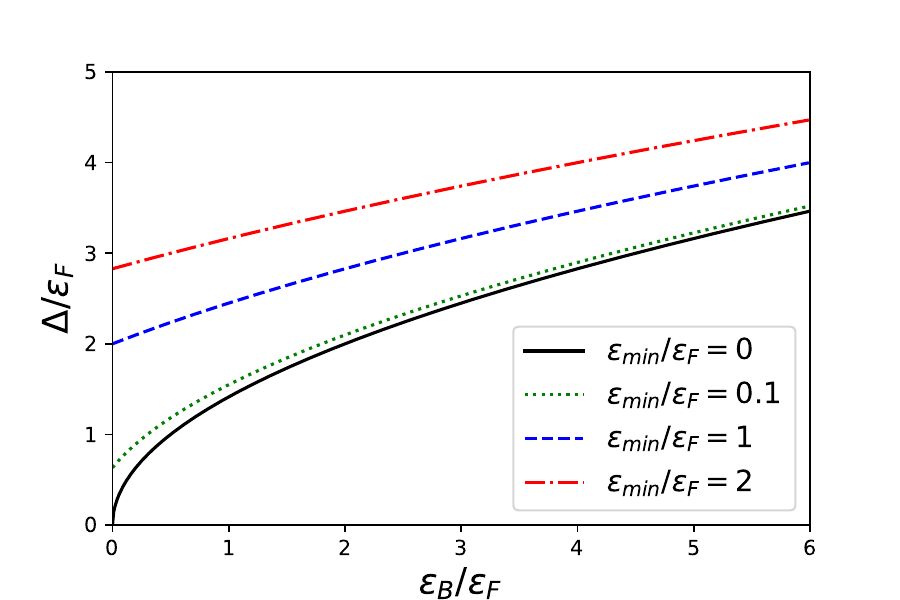}
    \small
  \caption{{\cblue On the left panel we have $\mu/\epsilon_{F}$ vs $\epsilon_{B}/\epsilon_{F}$ for different values of $\epsilon_{min}/\epsilon_{F}$, while on the right panel we picture $\Delta/\epsilon_{F}$ vs $\epsilon_{B}/\epsilon_{F}$ for different values of $\epsilon_{min}/\epsilon_{F}$. \newline In particular $\epsilon_{min}/\epsilon_{F}=$0 (black solid line), 0.1 (green dotted line), 1 (blue dashed line) and 2 (red dashed-dotted line).}}
  \label{cf_delta_MF}
\end{figure}

We investigate the behaviour of $\mu/\epsilon_{F}$ and $\Delta/\epsilon_{F}$ at different values of $\epsilon_{B}/\epsilon_{F}$ (see Fig. \ref{cf_delta_MF}). \\
As we can see, $\Delta$ is influenced by the value of the cutoff. As the cutoff increases, and so the system size decreases, the trend of the function $\Delta / \epsilon_{F}$ remains the same, but it moves towards negative x-axis. However, since negative values of the binding energy are not physical, we picture only the first quarter of the Cartesian plane.\\
The chemical potential $\mu$, instead, is not affected by changing
the system's size. 

Let us now consider the superfluid density. At the mean field level
it takes the following form
 \begin{equation}
 \label{ns0}
 n_s=n-n_{sp}
 \end{equation}
 where $n$ is the total density and $n_{sp}$ is the single particle contribution to the normal density. Following the
 Landau approach \cite{Landau_superfluidity} $n_{sp}$ can be written as
 \begin{widetext}
 \begin{equation}
\label{nn}
n_{sp}=\beta\int_{k_{min}}^\infty\frac{d^2\bf{k}}{(2\pi)^2}\frac{\hbar^2 k^2}{m}\frac{\exp\left[\beta\sqrt{(\frac{\hbar^2k^2}{2m}-\mu)^2+\Delta^2}\right]}{\left(\exp\left[\beta\sqrt{(\frac{\hbar^2k^2}{2m}-\mu)^2+\Delta^2}\right]+1\right)^2}
 \end{equation}
 \end{widetext}
which, after introducing $\epsilon =\hbar^{2} k^2/(2m)$, becomes
\begin{equation}
n_{sp}=\frac{m}{\hbar^2 \pi}\beta\int_{\epsilon_{min}}^\infty d\epsilon\, \epsilon \frac{\exp[\beta\sqrt{(\epsilon-\mu)^2+\Delta^2}]}{\left(\exp[\beta\sqrt{(\epsilon-\mu)^2+\Delta^2}]+1\right)^2}
\end{equation}
In the BCS limit, for $\mu>\epsilon_{min}$ and $\beta\gg 1$, the dependence on the cutoff $\epsilon_{min}$ is very weak and we can get the following approximation
\begin{equation}
\label{nn2}
n_{sp}\approx n\frac{\mu}{\epsilon_F}
\sqrt{{2\pi \beta \Delta}}\, \exp(-\beta\Delta)
\end{equation}
The finite-size dependence, in this limit, is, therefore, mainly due to $\Delta$, which, in the mean-field level, is given by eq.~(\ref{tildedeltakminneq0}). 
Increasing $\epsilon_{min}$, the gap $\Delta$ is enhanced, while $n_{sp}$ is suppressed. As a result, for finite size systems we expect that the superfluid density is promoted.

\section*{Appendix B: Number equation with infrared cutoff} 

The form for the gran potential in presence of Gaussian fluctuation is $\Omega=\Omega_{MF}+\Omega_{GF}$. The first term is the mean-field contribution while the last is due to Gaussian quantum fluctuations which reads
\begin{equation}
        \Omega_{\rm GF}=\frac{1}{2\beta} \sum_{\textbf{q},m} \ln{(\text{det } \mathbb{M}(\textbf{q}, \text{i}\Omega_{m}))}
\end{equation}
The matrix $\mathbb{M}(\textbf{q}, \text{i}\Omega_{m})$ is the inverse pair fluctuation propagator. It takes the form of a $2\times 2$ matrix, whose elements
are
\begin{widetext}
\begin{align}\label{first_matrix}
    M_{11} & (\textbf{q},\text{i}\Omega_{m}) = -\frac{1}{g} + \sum_{ \substack{ {\bf k} \\ |{\bf k}|\geq k_{min}}}
    \frac{\tanh{\left(\beta E_{\bf k}/2\right)}}{2 E_{\bf k}}  \times \nonumber \\& \times  \bigg[ \frac{(\text{i} \Omega_{m} - E_{\bf k} + \xi_{\bf k + q})(E_{\bf k} + \xi_{\bf k})}{(\text{i} \Omega_{m} - E_{\bf k} + E_{\textbf{k}+\textbf{q}})(\text{i} \Omega_{m} - E_{\bf k} - E_{\textbf{k}+\textbf{q}})} 
    -  \frac{(\text{i} \Omega_{m} + E_{\bf k} + \xi_{\bf k + q})(E_{\bf k} - \xi_{\bf k})}{(\text{i} \Omega_{m} + E_{\bf k} - E_{\textbf{k}+\textbf{q}})(\text{i} \Omega_{m} + E_{\bf k} + E_{\textbf{k}+\textbf{q}})}\bigg]
\end{align}
\begin{align}\label{second_matrix}
    M_{12} & (\textbf{q},\text{i}\Omega_{m}) = -
    \Delta^{2}\sum_{ \substack{ {\bf k} \\ |{\bf k}|\geq k_{min}}} \frac{\tanh{\left(\beta E_{\bf k}/2\right)}}{2 E_{\bf k}}  \times \nonumber \\& \times  \bigg[ \frac{1}{(\text{i} \Omega_{m} - E_{\bf k} + E_{\textbf{k}+\textbf{q}})(\text{i} \Omega_{m} - E_{\bf k} - E_{\textbf{k}+\textbf{q}})} + 
    \frac{1}{(\text{i} \Omega_{m} + E_{\bf k} - E_{\textbf{k}+\textbf{q}})(\text{i} \Omega_{m} + E_{\bf k} + E_{\textbf{k}+\textbf{q}})}\bigg]
\end{align}
\end{widetext}
where $\xi_{\bf k} = \hbar^{2}k^{2}/(2m) - \mu$ and $E_{\bf k} = \sqrt{\xi^{2}_{\bf k} + \Delta^{2}}$. Moreover, the following relations are valid
\begin{align}
    M_{11} (\textbf{q},-\text{i}\Omega_{m}) &= M_{22}(\textbf{q},\text{i}\Omega_{m}) \\
    M_{12} (\textbf{q},-\text{i}\Omega_{m}) &= M_{21}(\textbf{q},\text{i}\Omega_{m})   
\end{align}
The number equation can, then, be written as
\begin{equation}
    N = \frac{d\Omega}{d\mu} = \frac{\partial \Omega_{\text{MF}}}{\partial \mu} - \frac{\partial \Omega_{\text{GF}}}{\partial \mu} - \frac{\partial \Omega_{GF}}{\partial \Delta} \frac{\partial \Delta}{\partial \mu} \label{numb_eq_GF}
\end{equation}
From (\ref{numb_eq_GF}) we can get the chemical potential when we are in the presence of the Gaussian fluctuations. We calculate, following \cite{he2015quantum}, the energy density $E(\mu)$ and we look for its maximum value that gives the solution of $\mu$ for the given density $n$.
\\It is important to notice that the momentum integrals we compute in order to get $E(\mu)$ must be done also in the presence of the infrared cutoff in the lower bound. Moreover, in \cite{he2015quantum} a dimensionless parameter $\nu$ was introduced, which put in relation $\mu$ and $\epsilon_{B}$.  The presence of the infrared cutoff implies a new definition of $\nu$,
\begin{equation}
    \nu = \frac{\mu + \epsilon_{B}/2 + \epsilon_{min}}{\epsilon_{F}}
\end{equation}
This expression allows us, indeed, to obtain consistent results with the infinite-size case, both for the number and the gap equations. 

\section*{Appendix C: BKT Critical temperature} 
\begin{figure}[t]
    \centering
    \includegraphics[scale=.5]{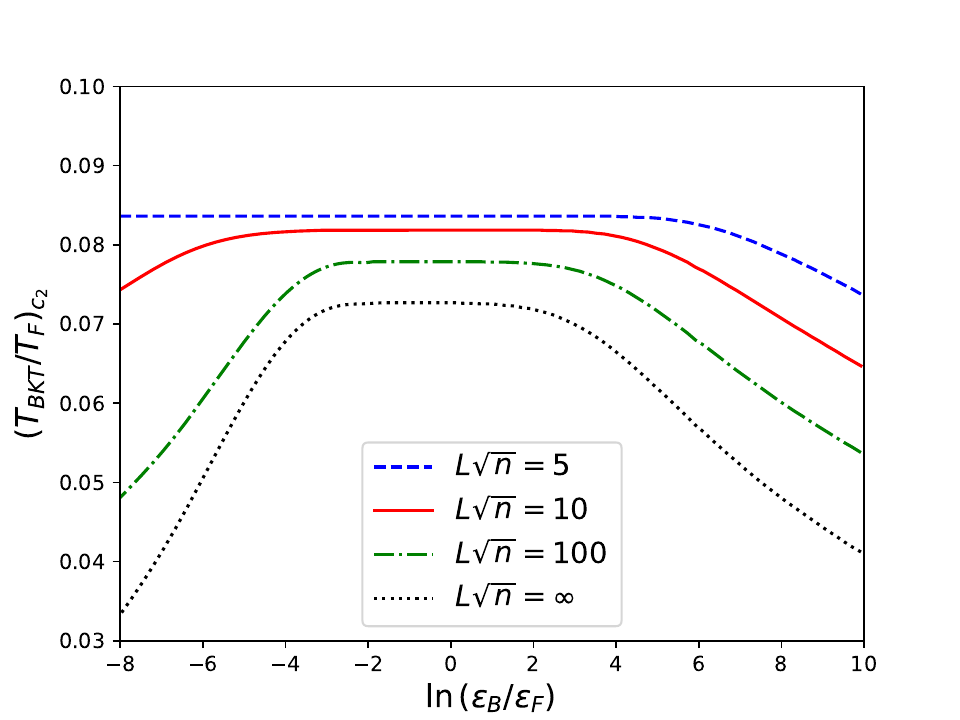}
    \caption{{\cblue Critical temperature obtained projecting the inflection points of the superfluid density to the temperature axis (see Fig. \ref{n_s_REN_cutoff_ln5} of the main text), 
    for $L\sqrt{n}$=5 (blue dashed line), 10 (red solid line), 100 (green dashed-dotted line) and $\infty$ (black dotted line).}}
    \label{fig:enter-label}
\end{figure}
As explained in the main text, we want to compute the critical temperature of the Berezinskii-Kosterlitz-Thouless phase transition. 
Since in a finite-size system we do not have a universal jump of the superfluid density $n_{s}$ we decided to propose two different procedures to compute $T_{BKT}$, based on the inflection points. In the first procedure we identify $T_{BKT}$
as the intersection of the tangent of $n_s$, as function of $T$, at the inflection point with the $T$-axis, for different values of the binding energy $\epsilon_B$ (the triangular points in Fig. \ref{n_s_REN_cutoff_ln5} of the main text). The results obtained by this procedure, $(T_{BKT/T_F})_{c_1}$, are reported in Fig. \ref{T_BKT_RG_cutoff_comparison} of the main text.  
The second criterion is identifying the critical temperatures just as the projections of the inflection points to the $T$-axis. By this second procedure we obtain the results for $T_{BKT}/T_F$ as functions of the binding energy, $(T_{BKT}/T_F)_{c_2}$ reported here in Fig. \ref{fig:enter-label}.

}

\end{document}